# Two-dimensional characterization of three-dimensional magnetic bubbles in Fe$_3$Sn$_2$ nanostructures

Jin Tang[1], Yaodong Wu[1,2], Lingyao Kong[3], Weiwei Wang[4], Yutao Chen[1], Yihao Wang[1], Y. Soh[5], Yimin Xiong[1], Mingliang Tian[1,3], and Haifeng Du[1,4*]

[1]Anhui Province Key Laboratory of Condensed Matter Physics at Extreme Conditions, High Magnetic Field Laboratory of the Chinese Academy of Sciences, and University of Science and Technology of China, Hefei, 230031, China

[2]Universities Joint Key Laboratory of Photoelectric Detection Science and Technology in Anhui Province, Hefei Normal University, Hefei, 230601, China

[3]School of Physics and Materials Science, Anhui University, Hefei, 230601, China

[4]Institutes of Physical Science and Information Technology, Anhui University, Hefei, 230601, China

[5]Paul Scherrer Institute, 5232, Villigen, Switzerland

*Corresponding author: duhf@hmfl.ac.cn

**ABSTRACT**

We report differential phase contrast scanning transmission electron microscopy (TEM) of nanoscale magnetic objects in Kagome ferromagnet Fe$_3$Sn$_2$ nanostructures. This technique can directly detect the deflection angle of a focused electron beam, thus allowing clear identification of the real magnetic structures of two magnetic objects including three-ring and complex arch-shaped vortices in Fe$_3$Sn$_2$ by Lorentz transmission electron microscopy imaging. Numerical calculations based on real material-specific parameters well reproduced the experimental results, showing that the magnetic objects can be attributed to integral magnetizations of two types of complex three-dimensional (3D) magnetic bubbles with depth-modulated spin twisting. Magnetic configurations obtained using the high-resolution TEM are generally considered as two-dimensional (2D) magnetic objects previously. Our results imply the importance of the integral magnetizations of underestimated 3D magnetic structures in 2D TEM magnetic characterizations.

# INTRODUCTION

Magnetic skyrmions are topologically non-trivial nanometric spin whirls that are expected to be information carriers in future energy-efficient spintronic devices [1-19]. They were first found in non-centrosymmetric magnetic compounds, where chiral Dzyaloshinskii-Moriya interactions (DMIs) bend the magnetic moments [20-23]. The unique feature of magnetic skyrmions is their non-trivial topology defined by unit topological charge [24]. Unlike the chiral DMI-induced skyrmions, magnetic bubbles originate from the interplay of four types of interactions including ferromagnetic exchange coupling, dipolar–dipolar interaction (DDI), uniaxial anisotropy, and Zeeman energy. Competition among the first three interactions leads to stripe domains, which may change into a magnetic bubble when applying an external field. There are two types of magnetic bubbles according to the rotation sense of the cylinder domain wall (Figure S1). One is a type-I magnetic bubble stabilized by a perpendicular magnetic field with a clockwise or anticlockwise closure cylinder domain wall contributing to a similar integer topological winding number as a skyrmion; type-I magnetic bubbles are thus renamed skyrmion bubbles [25-28]. The other one is a type-II magnetic bubble stabilized by a tilted magnetic field with magnetization in the partially reversed cylinder domain wall, with all domain wall magnetizations pointing toward the in-plane field component. However, such a domain wall in a type-II magnetic bubble contributes to a zero winding number and is topologically trivial [27]. The first wave of interest in magnetic bubbles occurred in the 1970s–1980s, motivated by experimental and theoretical studies of potential bubble memory [29,30]. The detection of skyrmion bubbles renewed the interest in magnetic bubbles in the last decade [25-28,31-35].

Although these two types of bubbles are well understood within the theoretical framework describing uniaxial ferromagnets, a recent study on a typical uniaxial ferromagnet $Fe_3Sn_2$ found new exotic spin whirls beyond conventional magnetic bubbles by Lorentz transmission electron microscopy (Lorentz-TEM) [25,28]. Two typical examples of such new spin whirls are three-ring and complex arch-shaped vortices characterized by a series of concentric circular stripe domains and one or multiple bound pairs of rotating magnetic whirls, respectively. Such magnetic structures were also observed in other uniaxial ferromagnets [26,31]. These objects are nanoscale size, which implies that they can be applied as information carriers in spintronic devices [17]. However, they are neither detected by other magnetic imaging methods nor in simulations conducted under realistic conditions. Moreover, a recent study demonstrated that the improper filter parameter in the transport of intensity equation (TIE) analysis of Lorentz-TEM imaging of type-II bubbles can lead to artificial biskyrmion structures [33].

Three-dimensional (3D) magnetic structures have become an active research topic because they are important in understanding novel experimental phenomena and potential applications [4,23,36-40]. It has been suggested that the chiral exchange interactions play important roles in tailoring 3D magnetic structures in synthetic antiferromagnets for potential 3D spintronic systems [39,40]. 3D magnetic skyrmions in $B_{20}$ magnets induced by DMI have been proposed to understand the stability of zero-field target skyrmions and attractive interactions between skyrmions [4,23]. Magnetic skyrmion bubbles have also been predicted with depth modulated spin twisting induced by DDI [41]. One typical characteristic of 3D magnetic skyrmion bubbles is that skyrmions near two surfaces have nearly contrary Néel-twisting. This characteristic has been observed in magnetic multilayers by some surface-sensitive magnetic detection methods [36-38]. TEM is a real-space imaging of integral magnetic field over depth with ultrahigh spatial resolution. Magnetic configurations in thin nanostructures have been typically considered as quasi two-dimensional (2D) magnetic objects using TEM [19,25,26,28,31]. However, one may clarify real 3D magnetic structures from the difference in integral magnetization over depth. This rule has been used to identify 3D chiral bobbers from integral phase shifts weaker than skyrmion tubes using TEM [3]. The depth-modulated 3D magnetic bubbles are also expected to show more complex integral magnetizations over the depth and are detected using 2D TEM magnetic imaging. The underestimated complex integral magnetizations of 3D magnetic bubbles may clarify the physics behind the complex three-ring and arch-shaped vortices in $Fe_3Sn_2$ through TEM, which is more readily considered as 2D magnetic configurations in thin nanostructures [25].

Here, we investigate the magnetic objects in an $Fe_3Sn_2$ nanodisc using differential phase contrast scanning transmission electron microscopy (DPC-STEM) combined with micromagnetic simulations. The observed magnetic objects are clarified as 2D integral magnetizations of complex 3D type-I and II bubbles with depth-modulated configurations. The characterization is considered further such that the origin of the artificial magnetic configurations detected in Lorentz-TEM is explained.

## RESULTS AND DISCUSSIONS
### Identification of a multi-ring vortex

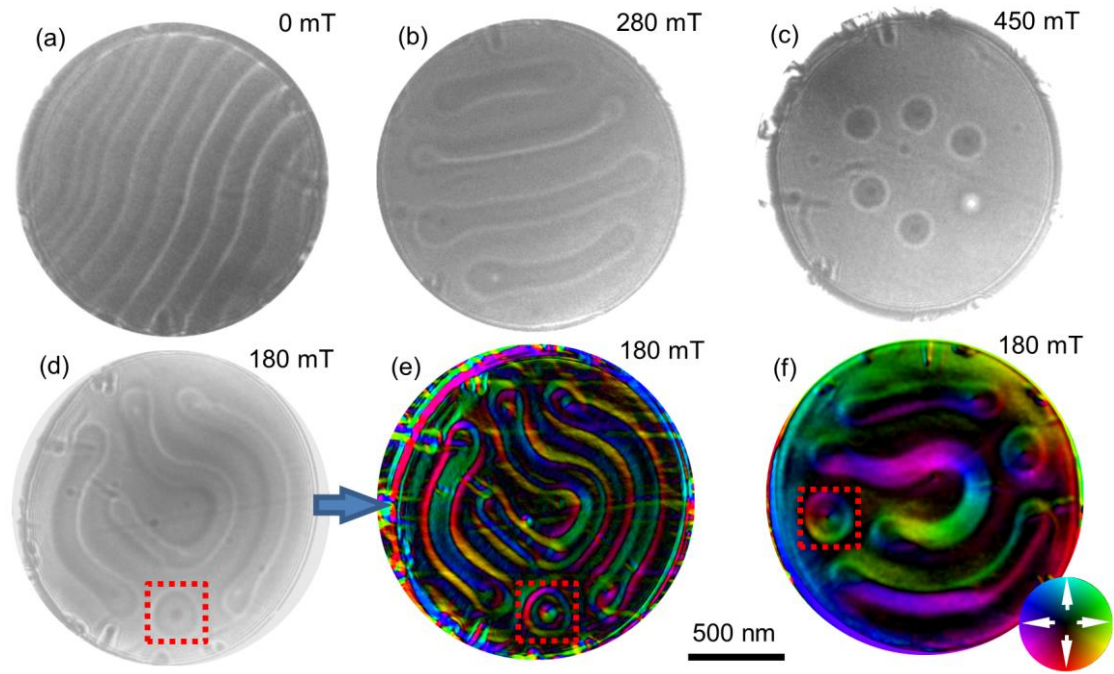

**Figure 1.** Magnetic field dependence of the spin configurations obtained using Lorentz-TEM at (a) 0, (b) 280, and (c) 450 mT. (d) Magnetic configuration obtained by decreasing the field from 450 to 180 mT. (e) The in-plane magnetic configuration from (d) reconstructed using TIE. A magnetic bubble marked by a red dot frame is chosen for the subsequent analysis in Figure 2. (f) DPC-STEM image of magnetic configuration at ca. 180 mT. The spin configurations in (e) obtained by Lorentz-TEM and (f) DPC-STEM are inconsistent because two magnetic imaging modes cannot be directly switched in our TEM setup. The color wheel represents the magnetization direction and amplitude; the dark area suggests the magnetization is out-of-plane.

We first focus on the three-ring vortex in $Fe_3Sn_2$ uniaxial ferromagnet. An $Fe_3Sn_2$ nanodisc (diameter, ca. 1550 nm; thickness, ca. 140 nm) with (001)-oriented out-of-plane direction is chosen for DPC-STEM measurements (Figure 1f and Figure S3) and micromagnetic simulations (see supplementary simulation method) [42]. Lorentz-TEM is also performed for comparison. TEM magnetic imaging is discussed in detail in supporting information [11-13,15,43-49]. Stripe domains are observed at zero field, which transfer into circular domains when a magnetic field is applied out-of-plane (Figures 1a–1c). However, once the circular domains are formed, they may persist as the field decreases (Figure 1d). In such a case, the Lorentz-TEM image gives rise to a three-ring vortex at low field (Figure 1e) that transfers into a normal bubble skyrmion when the field is increased. In Figure 2a, a field-driven process of one bubble by Lorentz-TEM is shown as an example. At a low field, a black dot in the center is surrounded by outer rings, which is different from a conventional skyrmion image [7,13,19]. The Lorentz contrast of a normal skyrmion is composed of only a black or white circle [5,6,19]. Such distinctness implies complexity in the magnetic objects. When using the TIE method, the reconstructed magnetic configuration is characterized by a series of concentric stripe domains with opposite rotation sense between neighboring magnetic rings (Figures 2 b1–b3), forming a three-ring vortex. At a high field, a normal skyrmion-like image is observed (Figures 2 b4–b5).

Assuming these nanoscale magnetic objects are arranged in thin nanostructures of uniform magnetization, such complex vortices with multiple rings and field-driven transition cannot be well reproduced in 2D uniaxial ferromagnets. However, we noted that the TEM method can only detect the integral in-plane magnetizations over the depth [5,6,19,45,46]. We noted the $Q$ factor of $Fe_3Sn_2$ determined by the ratio of uniaxial magnetic anisotropy (~54.5 $kJ/m^3$) to shape anisotropy (~244 $kJ/m^3$) is smaller than 1. In this case, DDI interaction could lead to the closure of cross-sectional bubble domains, which reveals Néel-twisting at the surface and Bloch-twisting in the middle [41]. Such Néel-twisting at two surfaces of 3D magnetic skyrmion bubbles with contrary chirality has been identified in reciprocal momentum space by a surface-sensitive resonant elastic X-ray scattering in magnetic multilayer films [8,36,37]. Using a nitrogen-vacancy magnetometer, a skyrmion in the surface layer has contrary chirality to intrinsic chiral interaction, which also implies the validity of the proposed 3D magnetic skyrmion bubbles [38]. Furthermore, more complex integral in-plane magnetizations over depth of 3D skyrmion bubbles that are measured using 2D TEM magnetic imaging will be expected and may explain the complex three-ring vortex (Figure 2b). We thus performed 3D numerical simulations of the $Fe_3Sn_2$ nanodisc, which showed field-driven evolutions of magnetic structures (Figure S4), similar to those observed in experiments (Figure 1). The main difference lies in the number of rotationally oriented magnetic rings at a low field. A two-ring vortex of simulated average in-plane magnetizations (Figures 2c) instead of three-ring vortex in Lorentz-TEM (Figures 2b) is obtained and characterized by a central weak vortex core and strong circular stripe domain around the edge. Simultaneously, the rotation sense of the outside ring and the central vortex are consistent and anticlockwise here. Such simulated results make sense intuitively because all the interactions in $Fe_3Sn_2$ are achiral. More importantly, such two-ring vortices in simulations (Figures 2c) are directly visualized by DPC-STEM (Figures 2d).

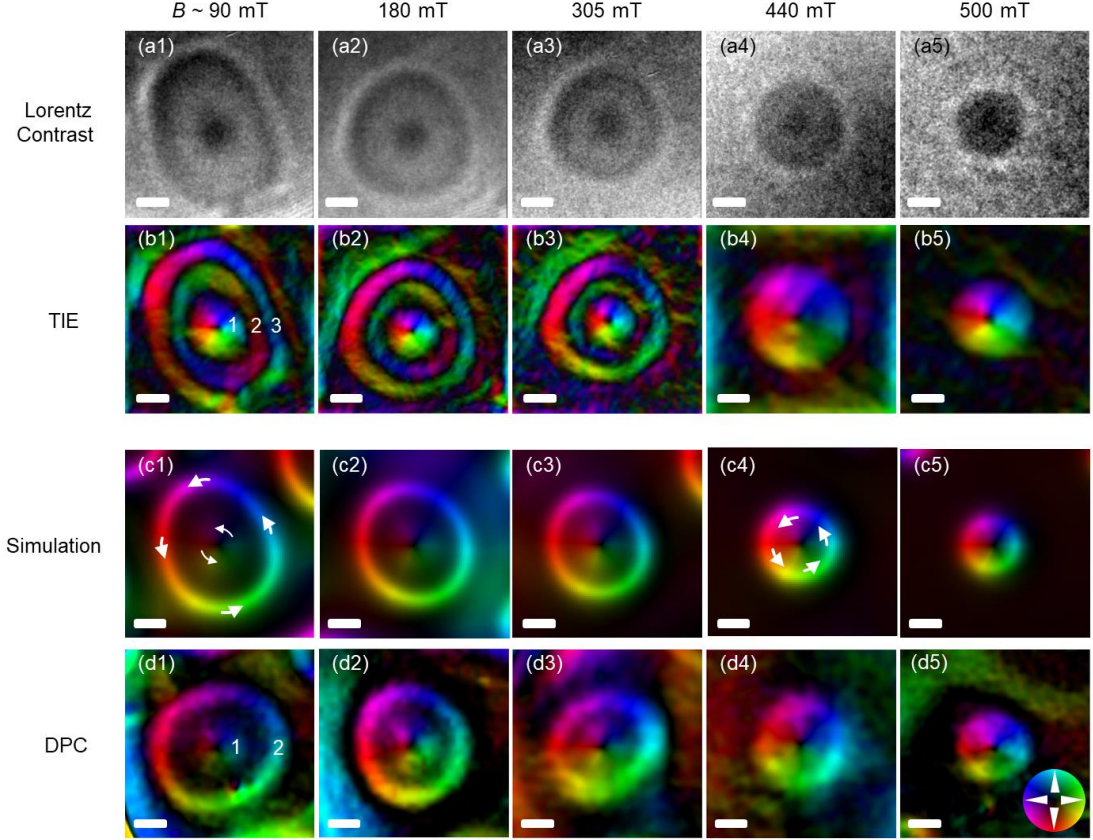

**Figure 2.** Variations of a magnetic bubble with field. (a1)–(a5) Intact magnetic contrast under defocused conditions in Lorentz-TEM; the defocus is 500 μm. (b1)–(b5) Magnetic configurations reconstructed by using TIE analysis. At low field (b1, 90 mT; b2, 180 mT; b3, 305 mT), a three-ring magnetic vortex is obtained; the ring number is marked in (b1). At high field (b4, 440 mT; b5, 500 mT), a normal skyrmion is obtained. $B$-dependence of the average in-plane magnetic configurations obtained by simulation (c1–c5) and DPC-STEM (d1–d5). The color wheel in (d5) indicates the direction and strength of the in-plane magnetization. Scale bar, 100 nm.

The consistency between the simulations and DPC-STEM imaging indicates an artifact in conventional Lorentz-TEM. A filter parameter $q_0$ is usually used in TIE to increase the signal-to-noise ratio of the reconstructed magnetic structure, avoid divergence, and suppress low-frequency disturbance represented by diffraction contrast, thus leading to deviation from the real features [33]. A clear transition from a two-ring magnetic vortex to the multiple-ring vortex with switched circulation is seen as $q_0$ increases (Figure S5). Such results imply that the other reported three-ring vortices from TIE analysis of Fresnel images that are not well understood should be re-examined using electronic holography or DPC-STEM to directly acquire the phase shift or phase gradient [26,31].

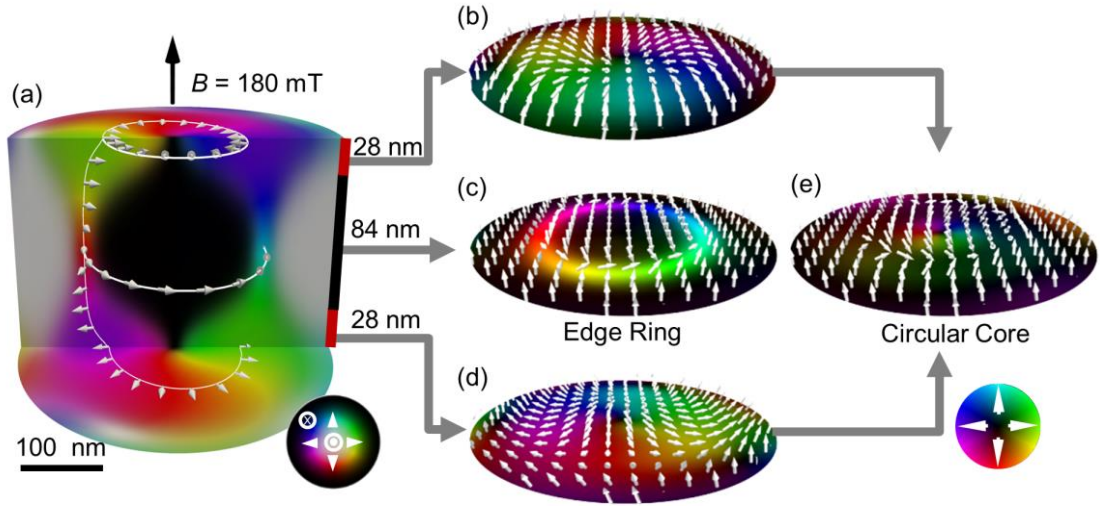

**Figure 3.** (a) Simulated 3D cross-section spin configurations of a two-ring vortex at 180 mT. (b) and (d) Average magnetic configuration around the upper and bottom surface at 28 nm depth. (e) Average magnetic configuration over the upper and bottom surfaces. (c) Average magnetic configuration around the center at a depth of 84 nm. The color wheels in (a) and (e) represent the in-plane magnetization orientation in (a) and (b)–(e), respectively. The white and darkness in the color wheel in (a) suggest the magnetization is out-of-plane up and down orientations, respectively.

The aforementioned consistency further enables us to analyze the origin of the two-ring vortex. The simulated 3D cross-section spin configuration of a two-ring vortex at a typical field is shown in Figure 3a. A rugby ball-like 3D structure is obtained, in which hybrid skyrmions along the sample thickness ranged from Néel- to Bloch-type with increasing depth below the surface, which is attributed to the DDI-induced vortex-like cross-sectional configurations. The surface layers host mainly Néel-type skyrmions with radially inward and outward pointing spins in the upper and bottom layers, respectively (Figures 3b and 3d). The Lorentz-TEM and DPC-STEM only detect the averaged in-plane magnetization, but much of the averaged in-plane magnetization cancels itself out, thus leading to a weak vortex core in the center (Figure 3e). From the 3D structure, it is readily understood that the outside ring originates from the Bloch-type skyrmions in the middle layers (Figure 3c), indicating that the two-ring vortex is intrinsically a type-I skyrmion bubble with depth-modulated spin configurations. Interestingly, when the field increased, the size of the outer ring, which comprises contributions from Bloch-type skyrmions in the middle layers, decreases from ca. 216 nm at $B \approx$ 90 mT to ca. 128 nm at $B \approx$ 450 mT. However, the size of the internal vortex-like core maintains constant (ca. 120 nm). Accordingly, at high field, the internal core and outer ring mix, leading to only one vortex (Figure S6), which may be responsible for traditional small-size one-ring skyrmion bubbles observed in Fe/Gd films with a comparable $Q$ factor as $Fe_3Sn_2$ [41].

Such agreement between the experimental and simulation results verifies the complex 3D structure of the type-I bubble skyrmion, which may give a general understanding of bubble skyrmions in uniaxial ferromagnets with a relatively small $Q$ factor [27,41]. We Noted that the presented two-ring vortices are distinct from the previously proposed two-ring bubbles in BaFeScMgO [31], which are typically target skyrmions with switched rotations and not attributed to the depth-modulated configurations.

# Identification of an arch-shaped vortex

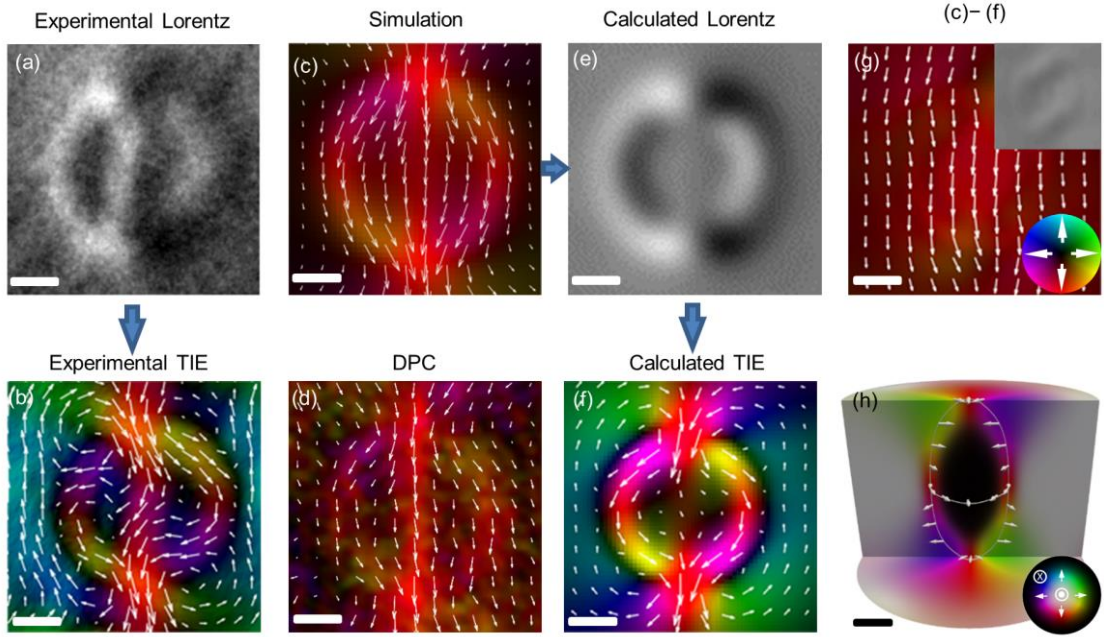

**Figure 4.** Magnetic configuration of an arch-shaped vortex. (a) The intact magnetic contrast in a defocused Fresnel image in Lorentz-TEM; defocus is 500 µm. (b) Magnetic configuration reconstructed using TIE with $q_0 = 0$. (c) Simulated averaged in-plane magnetization of a type-II magnetic bubble. (d) Representative DPC-STEM images of the magnetic configuration of a type-II bubble. (e) and (f) Calculated intact Lorentz contrast and the reconstructed magnetic configuration with $q_0 = 0$ based on the simulations in (c). (g) The difference between (f) and (c). (h) Simulated 3D cross-section spin configuration of the arch-shaped vortex with corresponding averaged in-plane magnetization shown in (c). The color wheels in (g) and (h) represent the in-plane magnetization amplitude and orientation in (b)–(g) and (h), respectively. Scale bar, 50 nm.

Following the procedure outlined previously to investigate the type-I bubble, here we discuss the type-II bubble to clarify the complex arch-shaped vortex [25,31]. According to our experiments, such a vortex can be easily obtained by slightly tilting the external field (Figure S7). The Lorentz contrast of such a vortex shows Φ-shaped ring with two strong contrasts on the top and bottom (Figure 4a). A weak line contrast in the center linking the two strong ones is also observed. Using the TIE method, the reconstructed magnetic configuration is characterized by multiple bound pairs of rotating magnetic whirls (Figure 4b).

The simulated averaged in-plane magnetic configuration (Figure 4c) shows a Φ-shaped spin whirl with the onion-like characteristic of a type-II bubble [27], which is confirmed using DPC-STEM images (Figure 4d). Based on the calculated magnetic configuration, the calculated Lorentz contrast (Figure 4e) is consistent with the experiments (Figure 4a), thus implying the correctness of the initial Lorentz contrast. However, the magnetic configuration reconstructed by TIE (Figure 4f) is entirely different from simulations and DPC-STEM images (Figures 4c and 4d). Therefore, we believe this magnetic object in Figures 4b and 4f is an artificial magnetic configuration created by TIE analysis.

    We compared the actual magnetic configuration and artificial magnetic configuration to obtain more insight into this issue. Interestingly, a nearly uniform ferromagnetic background is obtained if we subtract the magnetic configuration in Figure 4f from that in Figure 4c. Uniform magnetic configuration can only induce a uniform deflection of the electron beam. However, it cannot provide the Lorentz contrast (inset of Figure 4g) [50,51]. Therefore, there is no one-to-one correspondence between the Lorentz contrast and a real magnetic configuration. Generally, magnetic objects, differing by only a uniform ferromagnetic background, will exhibit the same Lorentz contrast. In a word, a ferromagnetic magnetization background is easily filtered out from the initial magnetization in the analysis of Lorentz-TEM contrast. We further show that the Φ-shaped spin whirl originates from a rugby ball-like 3D structure ranging from Néel- to Bloch-type with increasing depth below the surface (Figure 4h). The outside ring originates from the Bloch-type type-II bubble in the middle layers, and the central line comes from the averaged in-plane magnetization over the upper and bottom surfaces (Figure S8).

## CONCLUSIONS

In summary, using DPC-STEM magnetic imaging, we showed that 2D integral magnetizations of 3D type-I and II magnetic bubbles can well explain the multi-ring and arch-shaped vortices, respectively. The experimental observations are well reproduced by numerical calculations of real 3D magnetic nanostructures. We further analyzed the intrinsic origin of artifacts of magnetic contrast from Lorentz-TEM. Our results also imply that other unexplained magnetic configurations by TIE should be re-examined using other 2D TEM methods to consider their real 3D magnetic nanostructures [26,31]. In comparison to surficial magnetic configurations of 3D magnetic structures revealed by surface-sensitive methods [36-38], we provide a proof of the 3D magnetic bubbles in nanostructures from the view of 2D integral magnetizations. Given that the two types of bubbles are nanoscale magnetic objects, the next step is to study the dynamics induced by current to build a purely bubble-based spintronic device [28].

## METHODS

We prepared bulk $Fe_3Sn_2$ crystals by chemical vapor transport and fabricated the $Fe_3Sn_2$ nanodisc using a focused ion beam and scanning electron microscopy dual beam system (Helios NanoLab 600i, FEI). The magnetic imaging of the $Fe_3Sn_2$ nanodisc was performed on a TEM (Talos F200X, FEI) operated at 200 kV. Micromagnetic simulations were performed using a GPU-accelerated program: Mumax3. For details about the methods, refer to the supplemental information.

## Supporting Information

Supplementary data are available at *NSR* online.

## AUTHOR CONTRIBUTIONS

H.D. supervised the project. H.D. and J.T. conceived the experiments. Y-H.W. and Y.X. synthesized $Fe_3Sn_2$ single crystals. J.T. and Y.C. fabricated $Fe_3Sn_2$ nanodisks. J.T. and Y-D.W. performed TEM measurements. J.T. performed simulations. H.D., J.T., and L.K. prepared the first draft of the manuscript. All authors discussed the results and contributed to the manuscript.

## FUNDING

This work was supported by the National Key R&D Program of China, Grant No. 2017YFA0303201; the Key Research Program of Frontier Sciences, CAS, Grant No. QYZDB-SSW-SLH009; the Natural Science Foundation of China, Grants No. 11804343, 11974021, and U1432138; the President Foundation of Hefei Institutes of Physical Science, CAS, Grant No. YZJJ2018QN15; the Major/Innovative Program of Development Foundation of Hefei

Center for Physical Science and Technology, Grant No. 2016FXCX001; Universities Joint Key Laboratory of Photoelectric Detection Science and Technology in Anhui Province, Grant No. 2019GDTC06; and Anhui Province Key Laboratory of simulation and design for Electronic information system, Grant No. 2019ZDSYSZY04.

**Conflic of interstest statement**

None declared.

# Supporting Information:
# Two-dimensional characterization of three-dimensional magnetic bubbles in $Fe_3Sn_2$ nanostructures

Jin Tang[1], Yaodong Wu[1,2], Lingyao Kong[3], Weiwei Wang[4], Yutao Chen[1], Yihao Wang[1], Y. Soh[5], Yimin Xiong[1], Mingliang Tian[1,3], and Haifeng Du[1,4*]

[1]Anhui Province Key Laboratory of Condensed Matter Physics at Extreme Conditions, High Magnetic Field Laboratory of the Chinese Academy of Sciences, and University of Science and Technology of China, Hefei, 230031, China

[2]Universities Joint Key Laboratory of Photoelectric Detection Science and Technology in Anhui Province, Hefei Normal University, Hefei, 230601, China

[3]School of Physics and Materials Science, Anhui University, Hefei, 230601, China

[4]Institutes of Physical Science and Information Technology, Anhui University, Hefei, 230601, China

[5]Paul Scherrer Institute, 5232, Villigen, Switzerland

*Corresponding author: duhf@hmfl.ac.cn

# Methods

***Bulk sample preparations:*** Single Fe$_3$Sn$_2$ crystals were grown by chemical vapor transport with stoichiometric iron (Alfa Aesar, >99.9%) and tin (Alfa Aesar, >99.9%). The sintered Fe$_3$Sn$_2$ was obtained by heating the mixture at 800°C for 7 days, followed by thorough grinding. It was then sealed with I$_2$ in a quartz tube under vacuum and kept in a temperature gradient of 720°C to 650°C for 2 weeks. The crystal quality and structure group **R$\bar{3}$m** were both verified using Cu $K_\alpha$ radiation (TTR3 diffractometer, Rigaku).

***Fabrication of the Fe$_3$Sn$_2$ nanostructures:*** The thin Fe$_3$Sn$_2$ nanodisc used for TEM imaging had a diameter of 1550 nm and thickness of ca. 140 nm. It was fabricated by a lift-out process using a focused ion beam and scanning electron microscopy dual beam system (Helios NanoLab 600i, FEI) combined with a gas injection system and micromanipulator (OmniProbe 200+, Oxford). The nanodisc fabrication process was previously reported in detail [1,2].

***TEM measurements:*** High-resolution crystal structure and magnetic imaging were conducted in a TEM (Talos F200X, FEI) operated at 200 kV. Both Lorentz transmission electronic microscopy (Lorentz-TEM) and differential phase contrast (DPC) scanning TEM modes were used to obtain magnetic imaging, which can be qualitatively understood in terms of the Lorentz force expressed by **F** $=-e(\mathbf{v} \times \mathbf{B})$ acting on the electrons transmitted through a magnetic foil, where $e$, **v**, and **B** are the electric charge, the velocity of electrons, and magnetic field, respectively. Because the motion of the transmitted electrons is affected only by the magnetic induction perpendicular to the electron projection direction (out-of-plane), only the in-plane field ($B_{xy}$) can be detected using TEM in principle. The deflection angle $\beta_L$ of an electron induced by a magnetic field can be expressed as $\beta_L = B_{xy} e \lambda/h$ [3], where $\lambda$ is the wavelength of an elecrtron and $h$ is Planck's constant. Alternatively, based on the Aharonov–Bohm effect, the effect of the magnetic induction can be described as a phase shift $\varphi_M$ in quantum mechanics [4]. The phase gradient is derived to be proportional to the deflection angle and expressed as, $\nabla_{xy}\varphi_M = 2\pi\beta_L/\lambda$ [3]. The magnetic field $B$ that is proportional to the deflection angle $\beta_L$ can be obtained from the phase gradient $\nabla_{xy}\varphi_M$. The magnetic imaging using TEM is thus able to be realized by some special modes to acquire the phase shift $\varphi_M$. For varying field magnetic imaging, the objective lens of the microscope was turned off and adjusted to provide an out-of-plane magnetic field within a field ranging from −1700 to 1700 mT, as calibrated by a standard Hall probe. Traditional Lorentz-TEM acquires the phase shift $\varphi_M$ from the Fresnel images via a transport of intensity equation (TIE) process.

The DPC microscope was operated at low magnification in scanning TEM (STEM) mode using a split quadrant detector (Figure S2). The probe convergence and detection angles for the DPC-STEM measurements were set to 7 and 1 mrad, respectively; the corresponding probe size is ~3.6 nm. The beam deflection angle $\beta_L$ of the focused electron beam that is proportional to the phase gradient $\nabla_x \varphi_M$ is directly obtained from the intensity in each quadrant, named A, B, C, and D (Figure S2), of the segmented detector [3]. The orthogonal phase gradients along the *x* axis ($\nabla_x \varphi_M$) and the *y* axis ($\nabla_y \varphi_M$), are obtained by subtracting the signals of two orthogonal detectors, that is, differences (A − C) and (B − D), respectively. The phase gradients $\nabla_x \varphi_M$ and $\nabla_y \varphi_M$ are proportional to the magnetic field $-B_y$ and $-B_x$, respectively, by considering the phase shift induced by magnetic field according to Equation (12). The phase shift $\varphi_M$ is linear to an integral DPC (iDPC) image [5]. The strengths of the in-plane field (**B**$_{xy}$) were thus obtained through quadratic summation of the image differences ($\sqrt{(A-C)^2+(B-D)^2}$). The field orientation was determined by implementing the arctan of [(A–C)/(B–D)]. A typical analysis of DPC-STEM is shown in Figure S3. Note that our DPC-STEM setup cannot distinguish the phase gradient induced by electric and magnetic fields. The electric field reveals important information near a defect or the sample edge in our experiment and is thus ignored during the analysis of magnetic contrast.

***Micromagnetic simulations* [6]**: The micromagnetic simulations were performed using a GPU-accelerated program: Mumax$^3$. The total free energy terms are written as:

$$E=\int_{V_s} \{\varepsilon_{ex} + \varepsilon_{ani} + \varepsilon_{zeeman} + \varepsilon_{dem}\} d\mathbf{r}, \qquad (1)$$

where $\varepsilon_{ex}=A(\partial_x \mathbf{m}^2 + \partial_y \mathbf{m}^2 + \partial_z \mathbf{m}^2)$, $\varepsilon_{ani} = -K_u(\mathbf{u} \cdot \mathbf{m})^2$, $\varepsilon_{zeeman} = -M_s \mathbf{B} \cdot \mathbf{m}$, and $\varepsilon_{dem} = -\frac{1}{2} M_s \mathbf{B}_d \cdot \mathbf{m}$. Here $\mathbf{m} \equiv \mathbf{m}(x,y,z)$ is the normalized continuous vector field representing the magnetization $\mathbf{M} \equiv M_s \mathbf{m}(x,y,z)$, and $\mathbf{u}$ is a unit vector of uniaxial magnetic anisotropy. $A$, $K_u$, and $M_s$ are the exchange interaction, uniaxial magnetic anisotropy constant, and saturation magnetization, respectively. $\mathbf{B_d}$ is the demagnetizing field. The saturation magnetization $M_s$ of Fe$_3$Sn$_2$ is obtained from the magnetization of bulk Fe$_3$Sn$_2$ at a high field (7 T) at 300 K. The uniaxial magnetic anisotropy constant was determined from the saturation field ($H_k \approx 171$ mT) of the bulk Fe$_3$Sn$_2$ along the *ab* plane at 300 K by a function $K_u = 1/2\mu_0 H_k M_s$. The exchange interaction constant $A$ was adjusted to fit the average stripe domain width (ca. 175 nm), as shown in Figures 1(a) and S2a. Here we set the material parameters $M_s$ = 622.7 kA m$^{-1}$, $A$ = 8.25 pJ m$^{-1}$, and $K_u$ = 54.5 kJ m$^{-3}$, which are all typical parameters of the Fe$_3$Sn$_2$ uniaxial ferromagnet at room temperature. The cell size is set as 4 × 4 × 4 nm$^3$. The equilibrium spin configuration was obtained by using a conjugate gradient method.

***Simulation of the Lorentz-TEM image*** [7,8]: We obtained the configurations of the magnetic phase shift images based on the Aharonov–Bohm equation. The magnetic phase shift for an electron traversing the *z* axis is [9]:

$$\varphi_M(x,y) = -2\pi \frac{e}{h} \int A_z(\mathbf{r}) dz, \quad (2)$$

where, $A_z(\mathbf{r})$ is the magnetic vector potential. Magnetic vector potential has the contributions of demagnetization field $\mathbf{H_d}$ with an expression as $\nabla \times \mathbf{A} = \mu_0(\mathbf{M} + \mathbf{H_d})$. For a magnetic object with magnetization $\mathbf{M}(\mathbf{r})$, the magnetic vector potential is derived from the classic electrodynamics and expressed as [9]:

$$A_z(\mathbf{r}) = \frac{\mu_0}{4\pi} \int \mathbf{M}(\mathbf{r}') \times \frac{\mathbf{r}-\mathbf{r}'}{|\mathbf{r}-\mathbf{r}'|^3} d\mathbf{r}' \quad (3)$$

Combing the Eqations (2) and (3), we can obtained the magnetic phase shift and expressed as follows [8,9]:

$$\varphi_M(x,y) = -\mu_0 \frac{e}{h} \int \frac{(y-y')M_x(x',y') - (x-x')M_y(x',y')}{(x-x')^2 + (y-y')^2} dx'dy', \quad (4)$$

where $M_x$ and $M_y$ are the in-plane magnetization components that have been averaged over the *z* axis. They are expressed by: $M_x(x',y') = \int M_x(\mathbf{r}')dz'$, and $M_y(x',y') = \int M_y(\mathbf{r}')dz'$, respectively. We performed an integration over the thickness of the material along the line defined as $x = x'$ or $y = y'$. When the electrons have passed through the structure, thus acquiring a magnetic phase, they reach the back focal plane of the objective lens. Here, for simplicity, we only considered the contribution of the magnetic phase to the Lorentz-TEM image. The electron disturbance can be computed by performing a Fourier transform on the wave function of the transmitted electron beam:

$$g(k_x, k_y) = \iint \varphi_M(x, y) \exp[-2\pi i(k_x x + k_y y)] \mathrm{d}x \mathrm{d}y. \qquad (5)$$

The electron wave function is modified to $g(k_x, k_y) \, t(k_x, k_y)$ using a transfer function:

$$t(k_x, k_y) = A(k_x, k_y) \exp\left\{-2\pi i \left[\frac{C_s \lambda^3 k^4}{4} - \frac{\Delta f \lambda k^2}{2}\right]\right\}, \qquad (6)$$

where $\lambda$, $C_s$, and $\Delta f$ are the relativistic wavelength of the electrons, aberration coefficient of the objective lens, and defocus distance, respectively, and $k$ is expressed by $\sqrt{k_x^2 + k_y^2}$. We assumed the pupil function $A(k_x, k_y)$ was constant for all reciprocal space. In actual experiments, the Lorentz-TEM resolution is affected by the spread of the electron source and spatial coherence, which must be considered. We used an envelope function to describe the spread of the source, which is expressed as a Gaussian distribution:

$$E_s(k_x, k_y) = \exp\left[-\left(\frac{\pi\alpha}{\lambda}\right)^2 \left(C_s \lambda^3 k^3 + \Delta f \lambda k\right)^2\right]. \qquad (7)$$

Here, $\alpha$ is the beam divergence angle. The Lorentz-TEM intensity at the screen can be finally obtained as:

$$I(x', y') = \left|\iint g(k_x, k_y) t(k_x, k_y) E_s(k_x, k_y) \exp[-2\pi i(k_x x' + k_y y')] \mathrm{d}k_x \mathrm{d}k_y\right|^2. \qquad (8)$$

We set $\lambda$ = 0.0025 nm, thus corresponding to 200 keV electrons, and $C_s$ = 0 nm during the simulations. The beam divergence $\alpha$ is set as 0 mrad if it is not specified.

***Spin configuration obtained from magnetic TIE analysis* [10-12]**: Three Lorenz-TEM intensity contrast images at different defocus values (de-, in-, and over-focus) were analyzed by using the TIE analysis, and the phase shift can be obtained based on:

$$\varphi_M(x,y) = -\frac{2\pi}{\lambda} \nabla_{xy}^{-2} \nabla_{xy} \cdot \left[\left(\frac{1}{I(x,y,\Delta f)}\right) \nabla_{xy} \nabla_{xy}^{-2} \frac{\partial}{\partial \Delta f} I(x,y,\Delta f)\right]. \qquad (9)$$

We further used a Fourier transition and replaced the inverse Laplacian $\nabla_{xy}^{-2}$ by:

$$\nabla_{xy}^{-2} f(x,y) = \mathcal{F}^{-1}\left[\frac{\mathcal{F}[f(x,y)]}{|q(x,y)|^2}\right]. \tag{10}$$

Here, $q(x, y)$ is the spatial frequency in the image plane. Typically, a filter parameter $q_0$ is used to avoid divergence and suppress low-frequency noise represented by diffraction contrast, that is, to increase the signal-to-noise ratio of the obtained magnetic structure. Equation (8) must be replaced by:

$$\nabla_{xy}^{-2} f(x,y) = \mathcal{F}^{-1}\left[\frac{\mathcal{F}[f(x,y)]}{|q(x,y)|^2 + q_0^2}\right]. \tag{11}$$

Finally, we reconstructed the in-plane magnetic field from the following expression:

$$\mathbf{B} \times \mathbf{n} = -\frac{\hbar}{et} \nabla_{xy} \varphi_M(x,y), \tag{12}$$

where, $\hbar$ and $t$ are the reduced Planck constant and material thickness, respectively, and $\mathbf{n}$ is the unit vector parallel to the beam direction. The influence of the filter parameter $q_0$ on the retrieved magnetic structure is shown in Figure S5.

***Analyzing spin configurations in the micromagnetic simulations, simulated Lorentz-TEM images, and TIE analysis:*** We first obtained an initial equilibrium spin configuration in micromagnetic simulations using a conjugate gradient method. We then obtained the corresponding phase shift based on Equation (4). We further obtained the simulated Lorentz-TEM images under de-, in-, and over-focus conditions from the phase shift based on Equation (8). The corresponding phase shift can be further inversely obtained using the typical TIE analysis from the simulated Lorentz-TEM contrast images based on Equations (9)–(11). Finally, the magnetic field was reconstructed from the TIE analysis based on Equation (12).

# TEM MAGNETIC IMAGING

The basic principle of TEM magnetic imaging can be understood classically in terms of the deflection of the electron beams induced by the magnetic field **B**, as described in the Lorentz force law. Alternatively, a phase shift of an electronic wave can be induced by the magnetic field based on the Aharonov–Bohm effect in quantum mechanics [4]. It is thus derived that the phase shift gradient is proportional to the deflection [3]. Therefore, TEM magnetic imaging can be realized by some special modes to acquire a phase shift. Off-axis electronic holography, Lorentz-TEM (also called as in-line electron holography), and DPC-STEM are three main methods of TEM magnetic imaging [13]. Off-axis electronic holography directly acquires the phase shift from changes in the spacing of the interference fringes of the electron wave results from a biprism passing through a vacuum and magnetic field [13,14]. However, the sample size for electronic holography is limited to the effective width of the interference fringes for high-resolution magnetic imaging (typically ~1 μm for our TEM setting). Therefore, here we mainly use Lorentz-TEM and DPC for magnetic imaging of a 1550-nm $Fe_3Sn_2$ nanodisc.

The most commonly used technique in TEM magnetic imaging is the Fresnel method that obtains the phase shift using TIE analysis [10-13]. The Fresnel magnetic contrast originates from the superposition of two electron beams passing through two magnetic domains at out-of-focus conditions (see supplementary Figure S2a) [13]. As a result, Fresnel imaging in Lorentz-TEM is limited to domain walls, and uniform in-plane ferromagnetic magnetization leads to uniform Lorentz contrasts. Unlike the Lorentz-TEM, DPC-STEM relies on a quadrant-segmented detector that is directly used to obtain the deflection of a focused electron beam (Figure S2b). Accordingly, this technique directly detects the phase gradient along two orthogonal directions from two orthogonal subtracted detector signals (Figure S3). Furthermore, DPC-STEM is operated at an in-focus case, which avoids strong Fresnel fringes from defects or sample edges compared with Lorentz-TEM. Therefore, DPC-STEM has the advantages of in-focus STEM imaging and direct access to the phase gradient enable to visual domains and fine magnetic structures of skyrmions [15-17].

Other TEM imaging methods, such as electron tomography and ptychography, have been shown to have ultrahigh spatial resolution and even the ability to visualize 3D structures [14,18], but their applications on magnetic imaging are rare because of some unexplored issues. For example, electron tomography reconstructs 3D structures from a series of images obtained from a continuously rotating sample; such a process will introduce magnetic structure variations for a given out-of-plane field that renders magnetic structure reconstruction impossible. In the present case, electron tomography is expected to obtain zero-field 3D magnetic structures using field-free operations. Imaging 3D magnetic structures using electron tomography is expected to have wide applications in the developing rotatory field along with sample rotation.

For varying field magnetic imaging, the objective lens current was adjusted to provide an out-of-plane magnetic field, as calibrated by a standard Hall probe, in the Lorentz-TEM and DPC-STEM modes. DPC-STEM and Lorentz-TEM cannot be directly switched because the operation of switching magnetic imaging modes from Lorentz-TEM to DPC-STEM must be conducted under a demagnetizing field process from ~1700 mT in our present TEM setup.

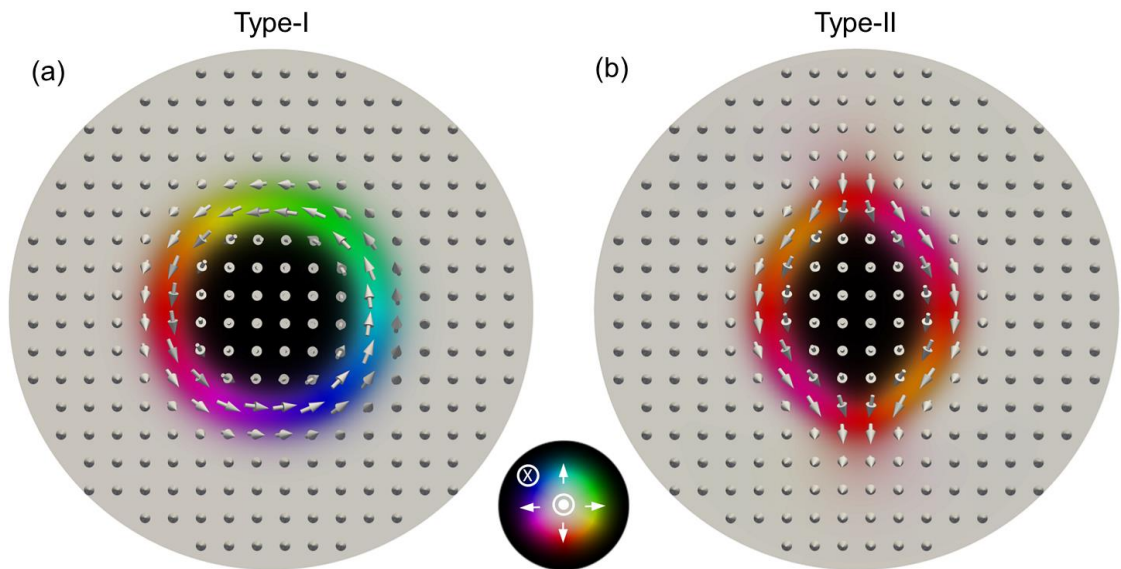

Figure S1. Representative schematic magnetic configurations for (a) a type-I bubble and (b) a type-II bubble. The color wheel indicates the direction of magnetization at each point; the white and darkness suggest magnetization point out-of-plane up and down orientation, respectively.

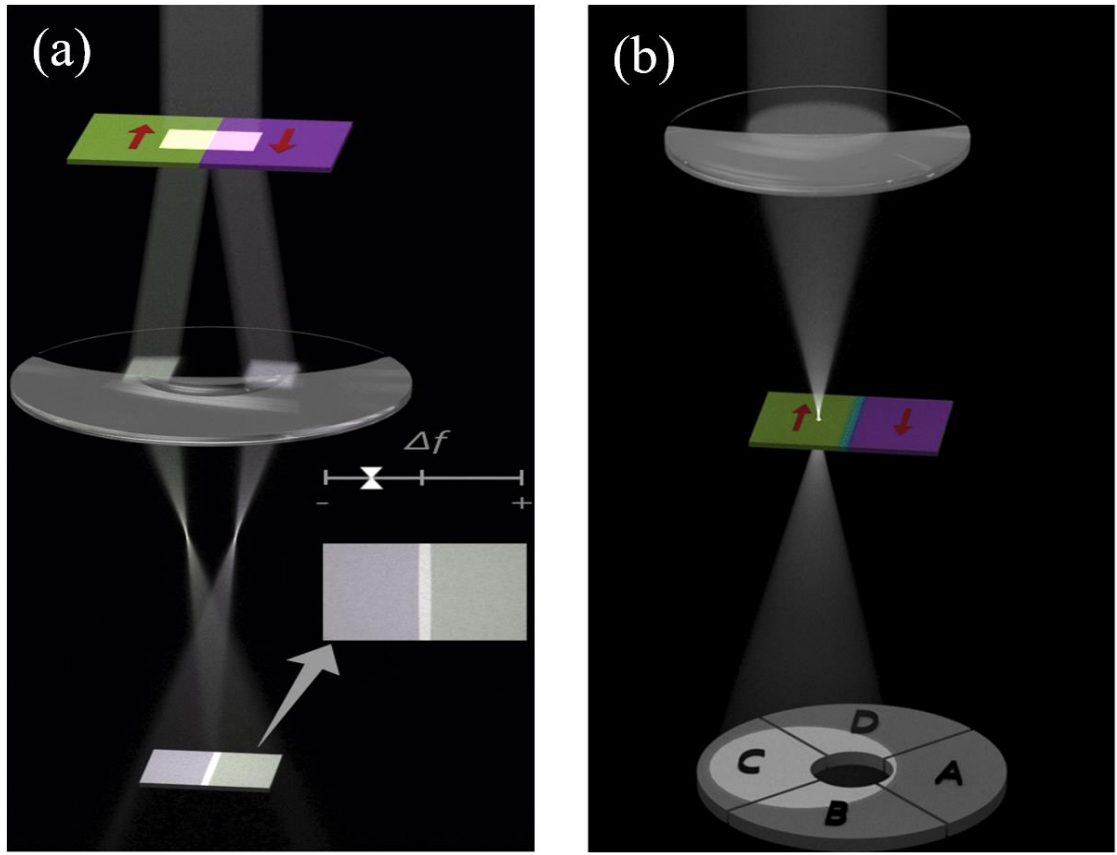

**Figure S2.** (a) Schematic ray diagram in a Fresnel image of a ferromagnetic domain specimen containing one 180° domain wall. A bright contrast around the domain wall forms by the superposition of two deflected electron beams. The appearance of dark or bright contrast depends on the out-of-focus conditions tuned by the focus distance $\Delta f$. (b) Schematic ray diagram in a DPC-STEM image. A beam of electrons is focused on a probe and scanned across the specimen. The beam deflected by the Lorentz force is collected using a detector with four segments. By considering the difference in the signals across the opposite quadrants, the direction and magnitude of the field can be directly deduced in real space, and the complete in-plane magnetic configurations can be mapped irrespective of the domain or domain wall.

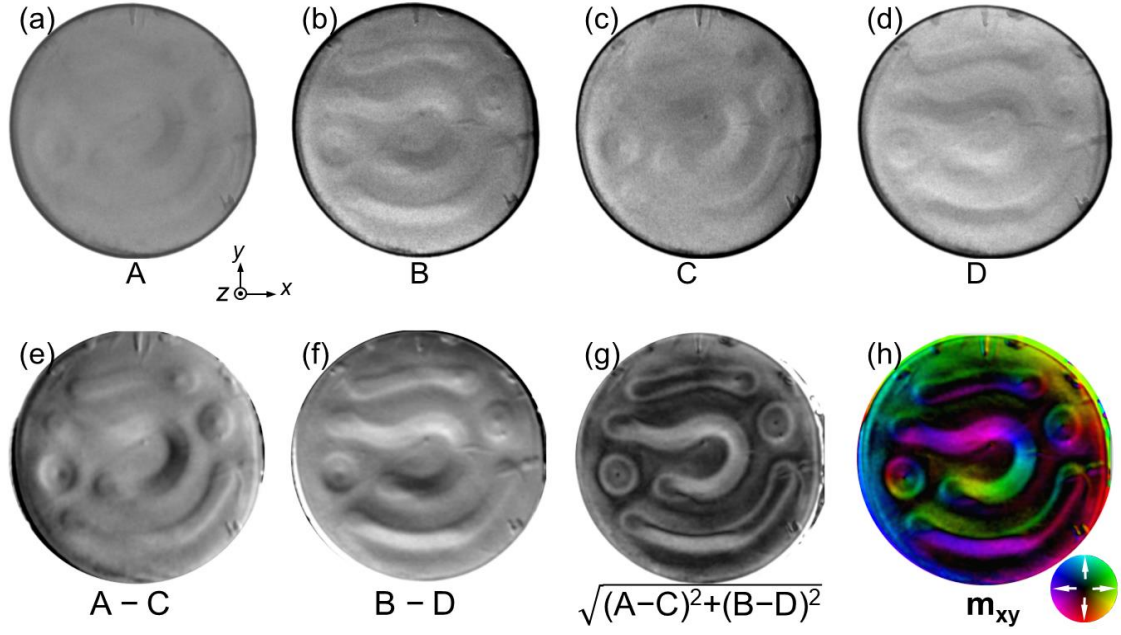

**Figure S3.** Analysis procedure for determining the magnetic structure in a 1550 nm Fe$_3$Sn$_2$ disc by using DPC-STEM. (a)–(d) DPC component images from the four segments of the detectors A, B, C, and D, respectively. (e) DPC component obtained by subtracting C from A (A–C) to detect the phase gradient along the $x$ axis $\nabla_x \varphi_M(x,y)$, which is proportional to the field component along the $y$ axis ($-\mathbf{B_y}$). (f) DPC component obtained by subtracting D from B (B–D) to detect the phase gradient along the $y$ axis $\nabla_y \varphi_M(x,y)$, which is proportional to the field component along the $x$ axis ($-\mathbf{B_x}$). (g) Total in-plane field strength ($|\mathbf{B_{xy}}|$) obtained from $\sqrt{(A-C)^2 + (B-D)^2}$. (h) Color mapping of in-plane field magnetization with direction and amplitude depicted based on the color wheel shown in the inset. The field direction is obtained by arctan[(A–C)/(B–D)]. The in-plane magnetization inside the magnetic sample ($\mathbf{m_{xy}}$) is proportional to the in-plane field ($\mathbf{B_{xy}}$).

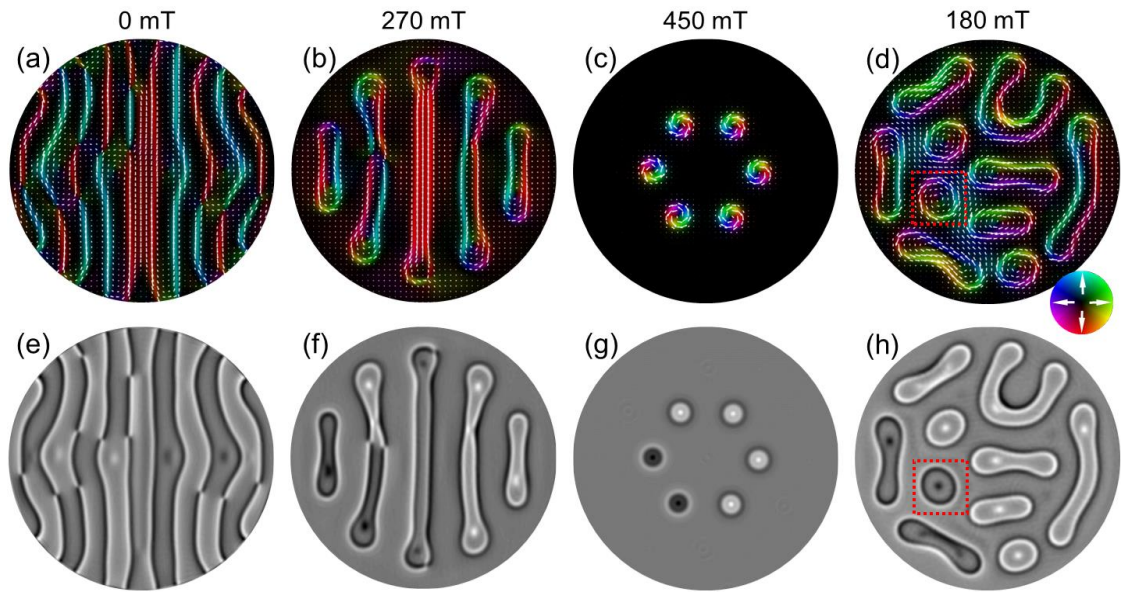

**Figure S4.** Simulated evolution of the magnetic structure in the 1550 nm ferromagnetic disc as a function of the external magnetic field. (a)–(c) Field-driven transition from initial stripe domains at the zero field to a bubble cluster at 450 mT. (d) Magnetic structure obtained by decreasing the magnetic field from 450 mT to 180 mT. (e)–(h) Corresponding simulated Lorentz contrast images at 0, 270, 450, and 180 mT taken under defocused conditions; defocus is 500 μm.

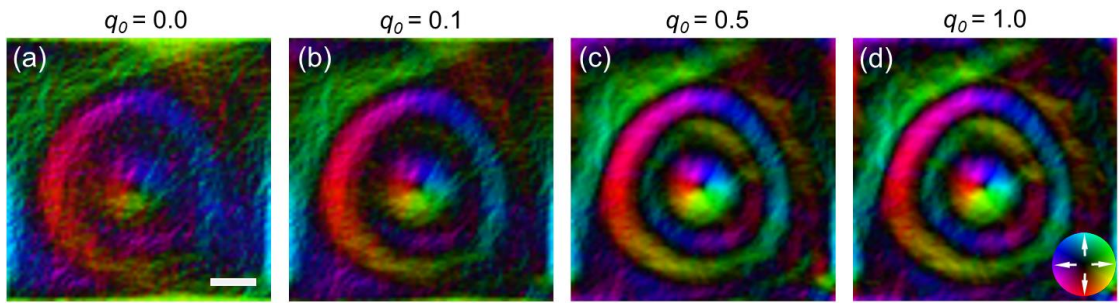

**Figure S5.** Reconstructed type-I bubble magnetic structure by TIE analysis with the filter parameter $q_0 = 0.0$ (a), $q_0 = 0.1$ (b), $q_0 = 0.5$ (c), and $q_0 = 1.0$ (d). The parameter $q_0$ significantly increases the signal-to-noise ratio of the retrieved magnetic structure. However, the magnetic structure transformed from a two-ring vortex into a three-ring vortex as $q_0$ increased. The scale bar is 100 nm.

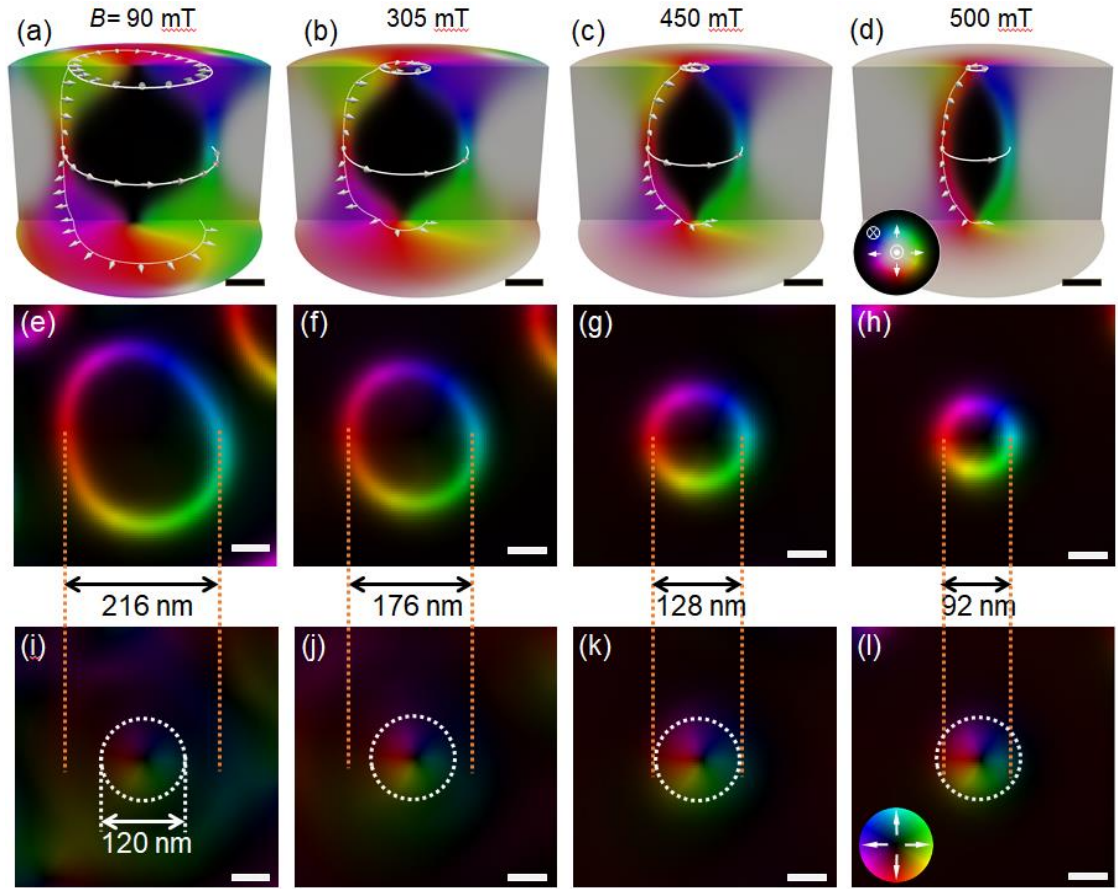

**Figure S6.** Field-driven evolution of a type-I bubble structure, as obtained by numerical simulation. Three-dimensional magnetic structures at (a) $B$ = 90, (b) 305, (c) 450, and (d) 500 mT. The in-plane magnetization mapping with amplitude and orientation in (a)–(d) are plotted according to the color wheel shown in the inset of (d). The averaged in-plane magnetization with contributions from the layers near the internal middle layers (84 nm) at (e) 90, (f) 305, (g) 450, and (h) 500 mT. The averaged in-plane magnetization with contributions from the layers near the surface (56 nm) at (i) 90, (j) 305, (k) 450, and (l) 500 mT. The in-plane magnetization mapping with amplitude and orientation in (d)–(l) are plotted according to the color wheel shown in the inset of (l). All scale bars are 50 nm. The diameters of the dotted circles in (i)–(l) are 120 nm.

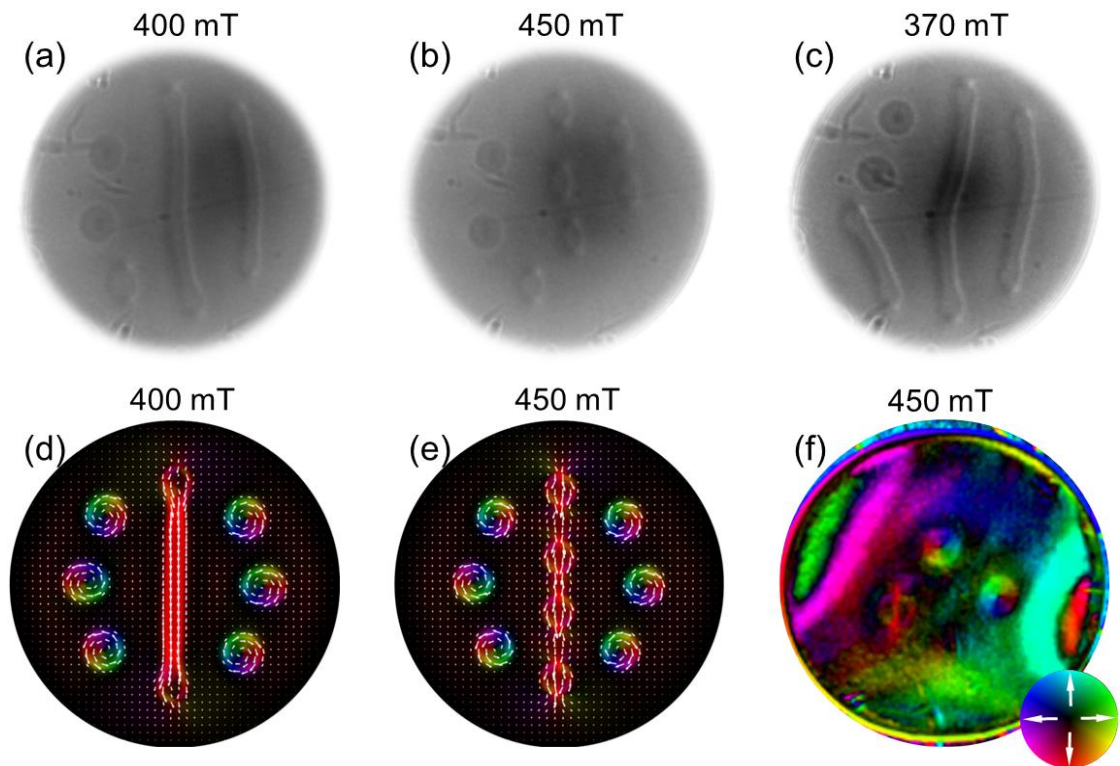

**Figure S7.** Transitions between the stripe domain and type-II bubble at a tilted magnetic field; angle is ca. 2.0 deg in the 1550 nm nanodisc. (a) Stripe domains with mixed bubbles at $B \approx 400$ mT by Lorentz-TEM. (b) Mixed type-I and type-II bubbles at $B \approx 450$ mT by Lorentz-TEM. (c) Transition from a type-II bubble to a stripe domain at $B \approx 370$ mT by Lorentz-TEM. (d)–(e) Simulated transition from a stripe domain at $B = 400$ mT to a type-II bubble at $B = 450$ mT. (f) Magnetic bubbles obtained by DPC-STEM at $B = 450$ mT.

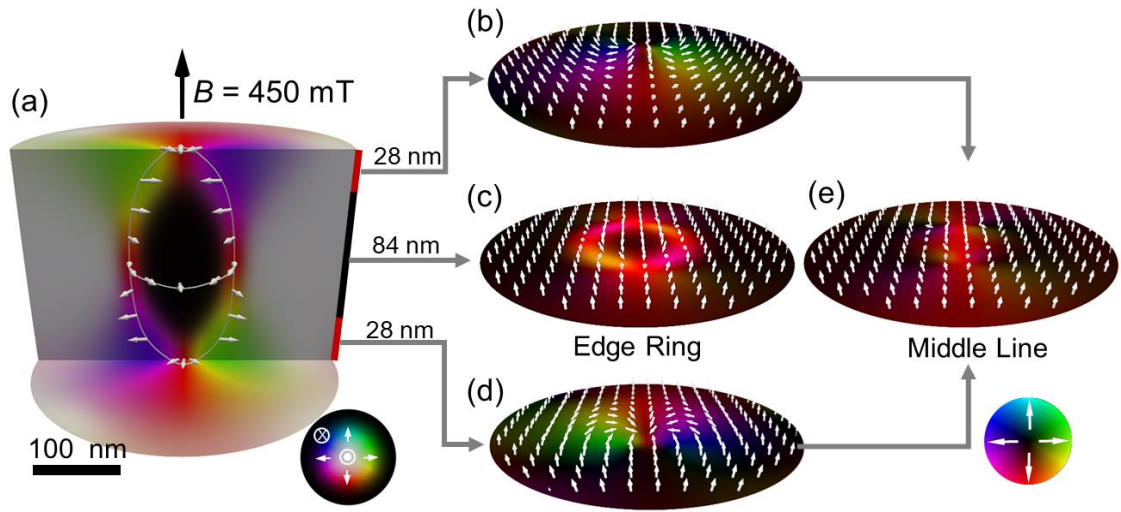

**Figure S8.** Three-dimensional depth-modulated magnetic structure of a type-II bubble. (a) Three-dimensional and cross-sectional spin configurations of a type-II bubble at 450 mT. Inset shows the color wheel representing the in-plane magnetization orientation and amplitude. (b) The averaged magnetization mapping in the top layers with 28-nm depth. (c) The averaged magnetization mapping in the middle layers with 84-nm depth revealing the edge ring. (d) The averaged magnetization mapping in the bottom layers with 28-nm depth. (f) Superposition of the top layers (b) and the bottom layers (d) revealing the middle line. The color mapping represents the in-plane orientation and amplitude shown in (b)–(e) are plotted according to the right color wheel.